\documentclass[twocolumn,showpacs,preprintnumbers,amsmath,amssymb,superscriptaddress,aps]{revtex4}
\usepackage{graphicx}% Include figure files
\usepackage{dcolumn}% Align table columns on decimal point
\usepackage{bm}% bold math
\usepackage{amsfonts}
\usepackage{mathrsfs}
\usepackage{graphicx}                   %for inserting figure
\usepackage{dcolumn}% Align table columns on decimal point
\usepackage{bm}% bold math
\usepackage{amsmath}

\begin{document}

%\preprint{YITP}

\title{Detection of quantum geometric tensor by nonlinear optical response}% Force line breaks with \\

\author{Zhi Li}
\email[Email:]{zhili@njust.edu.cn}
\affiliation{MIIT key Laboratory of Advanced Display Materials and Devices, Ministry of Industry and Information Technology, Institute of optoelectronics $\&$ Nanomaterials, Nanjing University of Science and Technology, Nanjing, 210094, China}
\author{Takami Tohyama}
\affiliation{Department of Applied Physics, Tokyo University of Science, Katsushika, Tokyo 125-8585, Japan}
\author{Toshiaki Iitaka}
\affiliation{RIKEN Center for Computational Science, RIKEN, 2-1 Hirosawa, Wako, Saitama 351-0198, Japan}
\author{Haibin Su}
%\email[Email:]{haibinsu@ust.hk}
\affiliation{Department of Chemistry, The Hong Kong University of Science and Technology, Hong Kong, China}
\author{Haibo Zeng}
%\email[Email:]{haibo@njust.edu.cn}
\affiliation{MIIT key Laboratory of Advanced Display Materials and Devices, Ministry of Industry and Information Technology, Institute of optoelectronics $\&$ Nanomaterials, Nanjing University of Science and Technology, Nanjing, 210094, China}
%\author{Haibin Su}
%\email[Email:]{haibinsu@ust.hk}
%\affiliation{Department of Chemistry, The Hong Kong University of Science and Technology, Hong Kong, China}
%\author{Hongming Weng}
%\email[Email:]{hmweng@iphy.ac.cn}
%\affiliation{Beijing National Laboratory for Condensed Matter Physics, and Institute of Physics, Chinese Academy of Sciences, Beijing 100190, China}
%\affiliation{Beijing National Laboratory for Condensed Matter Physics, and Institute of Physics, Chinese Academy of Sciences, Beijing 100190, China}
%\affiliation{Institute of Advanced Studies, Nanyang Technological University, 60 Nanyang View, 639673 Singapore}

\date{\today}% It is always \today, today,

\begin{abstract}
 Quantum geometric tensor (QGT), including a symmetric real part defined as quantum metric and an antisymmetric part defined as Berry  curvature, is essential for understanding many phenomena. We studied the photogalvanic effect of a multiple-band system with time-reversal-invariant symmetry by theoretical analysis in this work. We concluded that the integral of gradient of the symmetric part of QGT in momentum space is related to the linearly photogalvanic effect, while the integral of gradient of Berry curvature is related to the circularly photogalvanic effect. Our work afforded an alternative interpretation for the photogalvanic effect in the view of QGT, and a simple approach to detect the QGT by nonlinear optical response.
\end{abstract}
%\pacs{73.43.-f, 71.20.Mq, 73.61.Ey}

\maketitle

%\tableofcontents
%\section{Introduction}
    \emph{Introduction}  Geometry plays an important role in many aspects of modern physics\cite{Xiao10,Resta94,Resta07,AP17}. The geometry of the eigenstates is encoded in the quantum geometric tensor (QGT)\cite{GV80,Berry89,WML10}, which is defined on any manifold of states smoothly varying with some parameter $\lambda$. The geometric tensor naturally appears when one defines the 'distance' between nearby states $|\psi(\lambda)\rangle$ and $|\psi(\lambda+d\lambda)\rangle$
    \begin{equation}\label{ds}
    \begin{split}
      ds^{2}=1-|\langle\psi(\lambda)|\psi(\lambda+d\lambda)\rangle|^{2}   \\
      =d\lambda_{\alpha}Q_{\alpha\beta}d\lambda_{\beta}+O(|d\lambda|^{3})
    \end{split}
    \end{equation}
    where $Q_{\alpha\beta}=\Gamma_{\alpha\beta}+i\Omega_{\alpha\beta}$ is as object known as the geometric tensor comprising the antisymmetric Berry curvature $\Omega_{\alpha\beta}$ and the symmetric quantum metric $\Gamma_{\alpha\beta}$. The Berry curvature is crucial for crucial for topological matter\cite{Kane10,XL11,Ando13,FD17,AV18,BAB16,WHM16}, and the quantum metric defines the distance between the eigenstates. Knowledge of the quantum metric is essential for understanding many phenomena, such as, orbital magnetic susceptibility\cite{QN14}, the exciton Lamb shift\cite{AI15}, and anomalous Hall drift\cite{GM20}.

    Nonlinear optical (NLO) response\cite{NLO1}, including second harmonic generation (SHG) and photogalvanic effect, has wide applications in scientific community\cite{Boyd}. For example, SHG is used for frequency doubling of laser light, and detection of the breaking of spatial inversion symmetry (SIS)\cite{Shen}. The photogalvanic effect (PGE), where the dc electronic current is induced when the material is illuminated by light, can occur in materials with broken space inversion symmetry\cite{Deyo09,TM16,EJK17,JEM19,MP16,at17}. There are two types of photogalvanic effect in semiconductors, i.e., injection current induced by circularly polarized light and shift current induced by linearly polarized light. The injection current is also dubbed as circular photogalvanic effect (CPGE) because the direction of current is determined by the helicity of light, and CPGE can be used for the detection of topological charge of quantum matter\cite{PH11,JO19,CPGE1}. The direction of shift current is independent from the helicity of light, and usually dubbed as linear photogalvanic effect (LPGE). Recently, the Berry curvature dipole (BCD) defined as the integral of the gradient of the Berry curvature in momentum space\cite{Fu15,DC17,DC19,IS19,Niu19}, affords a new interpretation for PGE. Bilayer WTe$_{2}$ with tilted Weyl point provides a possible platform to detect the BCD and its induced nonlinear Hall current\cite{HZ18,NLO19,RN19}. The nonlinear Hall effect induced by BCD implies that the nonlinear response may has some underlying connection with the QGT. We may wonder that what is the role of symmetric real part of QGT for nonlinear optical response, and can we detect QGT by nonlinear optical response?

    In this letter, we address the basic theory of GPE of multiple-band system with time-reversal-invariant symmetry (TRIS). We concluded that the gradient of symmetric real part of QGT is related to the LPGE, while the gradient of Berry curvature is related to the CPGE. Our work builds close connections between nonlinear optical response of a system with time-reversal-invariant symmetry and the geometry of quantum states, and facilitates the detection of QGT by nonlinear optical response.

%\section{quantum kinetic equation}

    \emph{Dynamics of density matrix}  Under spatial homogeneous external field $\vec{E}(t)=\int_{0}^{\infty} d\omega[\vec{E}(\omega)e^{-i\omega t}+c.c.]$, the light-matter interaction can be described by below model Hamiltonian,
    \begin{equation}\label{H}
    \hat{H}(t)=\int d\vec{r}\Psi^{\dagger}(\vec{r},t)[\hat{h}_{0}-e\vec{E}(t)\cdot \vec{r}]\Psi(\vec{r},t),
    \end{equation}
  where $\hat{h}_{0}$ is the unperturbed Hamiltonian describing the ground state. The orthogonal Bloch functions $\phi_{n}(\vec{k},\vec{r})$ satisfy
  \begin{equation}\label{H0}
    h_{0}\phi_{n}(\vec{k},\vec{r})=\epsilon_{n}(\vec{k})\phi_{n}(\vec{k},\vec{r}),
    \end{equation}
    in which \emph{n} is the band index and $\vec{k}$ is position in momentum space. The Bloch functions are orthogonal to each other,
    \begin{equation}\label{O}
    \int d\vec{r}\phi_{n}^{\dagger}(\vec{k},\vec{r})\phi_{m}(\vec{k'},\vec{r})=\delta_{nm}\delta(\vec{k}-\vec{k}').
    \end{equation}
   Wave function $\Psi(\vec{r},t)$ can be expressed as combination of Bloch functions with annihilation operators $a_{n}(\vec{k})$,
      \begin{equation}\label{F}
      \begin{split}
    \Psi(\vec{r},t)=\sum_{n}\int_{BZ}\frac{d^{3}\vec{k}}{(2\pi)^{3}}a_{n}(\vec{k},t)\phi_{n}(\vec{k},\vec{r})   \\
    =\sum_{n,\vec{k}}a_{n}(\vec{k},t)\phi_{n}(\vec{k},\vec{r}).
    \end{split}
    \end{equation}
 The velocity operator\cite{Xiao10} is defined as
    \begin{equation}\label{V}
    \vec{v}=\frac{i}{\hbar}[\hat{H},\vec{r}]=\frac{i}{\hbar}[\hat{h}_{0},\vec{r}]-\frac{ie}{\hbar}[E_{\alpha}(t)\hat{x}^{\alpha},\vec{r}],
    \end{equation}
    where we assume that summation is performed on repeated index, and the current is expressed as
    \begin{equation}\label{J}
    \begin{split}
    J_{u}(t)=\frac{ie}{\hbar}\Psi^{\dagger}(t)[\hat{H},\hat{x}_{u}]\Psi(t)=  \\
    \frac{e}{\hbar}\int d\vec{k}\partial_{u}\epsilon_{n}(\vec{k})\rho_{nn}(\vec{k},t)+    \\
    \frac{ie}{\hbar}\int d\vec{k}\epsilon_{nm}(\vec{k})A_{nm}^{u}(\vec{k})\rho_{nm}(\vec{k},t)- \\
    \frac{e^{2}E_{v}(t)}{\hbar}\int d\vec{k}f_{nm}^{uv}(\vec{k})\rho_{nm}(\vec{k},t)+       \\
    \frac{e^{2}E_{v}(t)}{\hbar}\int d\vec{k}g_{n}^{uv}(\vec{k},t)
    \end{split}
    \end{equation}
    where $\partial_{u}=\frac{\partial}{\partial k_{u}}$, energy difference $\epsilon_{nm}(\vec{k})=\epsilon_{n}(\vec{k})-\epsilon_{m}(\vec{k})$, and density matrix $\rho_{nm}(\vec{k},t)=a_{n}^{\dagger}(\vec{k},t)a_{m}(\vec{k},t)$.
    Here, we used the position matrix\cite{B1962,ZL18,ZL19} and the definition of Berry connection,
   \begin{equation}\label{P}
   \begin{split}
    \langle n'\vec{k}'|\vec{r}|n\vec{k}\rangle=\int d^{3}\vec{r}\phi_{n'}(\vec{k'},\vec{r})\vec{r}\phi_{n}(\vec{k},\vec{r})=  \\
    A_{n'n}(\vec{k})\delta(\vec{k}-\vec{k}')-i\delta_{n'n}\nabla_{\vec{k}}\delta(\vec{k}-\vec{k}'),
   \end{split}
    \end{equation}
    where Berry connection $\vec{A}_{nm}(\vec{k})=i\langle u_{n}(\vec{k})|\nabla_{\vec{k}}u_{m}(\vec{k})$, in which $u_{n}(\vec{k})=e^{-i\vec{k}\cdot \vec{r}}\phi_{n}(\vec{k},\vec{r})$ is the periodic part of Bloch function. The (static) non-abelian Berry curvature $f_{mn}^{uv}(\vec{k})$ is defined as\cite{XL08},
    \begin{equation}\label{NAC}
    \begin{split}
    f_{nm}^{uv}(\vec{k})=\partial_{u}A_{nm}^{v}(\vec{k})-\partial_{v}A_{nm}^{u}(\vec{k})-  \\
    i\sum_{l}[A_{nl}^{u}(\vec{k})A_{lm}^{v}(\vec{k})-A_{nl}^{v}(\vec{k})A_{lm}^{u}(\vec{k})]
    \end{split}
    \end{equation}
    which is vanishing, and the dynamical Berry curvature $g_{n}^{uv}(\vec{k},t)$ is defined as,
     \begin{equation}\label{Dynamical}
     \begin{split}
    g_{n}^{uv}(\vec{k},t)=i[\partial_{u}a_{n}^{\dagger}(\vec{k},t)\partial_{v}a_{n}(\vec{k},t)-  \\
    \partial_{v}a_{n}^{\dagger}(\vec{k},t)\partial_{u}a_{n}(\vec{k},t)].
    \end{split}
    \end{equation}
    TRIS required that $g_{n}^{uv}(\vec{k},t)$=$-g_{n}^{uv}(-\vec{k},t)$, therefore the dynamical Berry curvature will not cause GPE. Therefore, only the second line and third line may contribute to the second order response.

    The dynamics of density matrix $\rho(\vec{k},t)$ can be described by the Heisenberg's equation of motion ~\cite{mp18},
  \begin{equation}\label{EOM}
   \begin{split}
    i\hbar\frac{d\rho_{nm}(\vec{k},t)}{\partial t}= i\hbar\frac{da_{n}^{\dagger}(\vec{k})}{dt}a_{m}(\vec{k})+i\hbar a_{n}^{\dagger}(\vec{k})\frac{da_{m}(\vec{k})}{dt}   \\
    =[a_{n}^{\dagger}(\vec{k}),\hat{H}]a_{m}(\vec{k})+a_{n}^{\dagger}(\vec{k})[a_{m}(\vec{k}),\hat{H}].
     \end{split}
  \end{equation}
 For the intrinsic NLO effect from geometry of eigenstates, we ignore all the electron-electron, electron-phonon, and electron-impurity scattering term here. We expand the density matrix up to the second order of field strength,
    \begin{equation}\label{Expansion}
    \rho_{nm}(\vec{k},t)=\rho^{(0)}_{nm}(\vec{k})+\rho^{(1)}_{nm}(\vec{k},t)+\rho^{(2)}_{nm}(\vec{k},t)+\ldots,
    \end{equation}
  in which $\rho^{(\lambda)}_{nm}(\vec{k},t)\propto |E|^{\lambda}$, and $\rho^{(0)}_{nm}(\vec{k})=\delta_{nm}\rho^{(0)}_{nn}(\vec{k})$ is the electronic distribution of ground state. Here, $\rho^{(0)}_{nn}=\frac{1}{1+\exp(\frac{\epsilon_{n}}{k_{B}T})}$ ($k_{B}$ Boltzmann constant, \emph{T} temperature) is fermi-Dirac distribution of band \emph{n}.
  From Eq.~(11), the first-order frequency dependent inter-band (n$\neq$m) density matrix reads,
  \begin{equation}\label{rho1}
  \rho^{(1)}_{nm}(\vec{k},\omega)=\frac{e\vec{E}(\omega)\cdot \vec{A}_{mn}(\vec{k})[\rho^{(0)}_{mm}(\vec{k})-\rho^{(0)}_{nn}(\vec{k})]}{\hbar\omega+\epsilon_{nm}(\vec{k})+i\eta},
  \end{equation}
  where $\eta$ is an infinitesimal parameter. The first-order intra-band frequency dependent density matrix reads
  \begin{equation}\label{rhointra}
  \rho^{(1)}_{nn}(\vec{k},\omega)=\frac{-ie}{\hbar\omega}\vec{E}(\omega)\cdot \partial_{\vec{k}}\rho^{(0)}_{nn}(\vec{k}).
  \end{equation}
  From Eq.~(11), the second-order frequency dependent intra-band and inter-band density matrices read,
   \begin{widetext}
   \begin{eqnarray}
    \nonumber 
     \rho_{nn}^{(2)}(\omega_{3} &=& \frac{-ie}{\hbar\omega_{3}}\sum_{\omega_{1},\omega_{2}}\delta(\omega_{3},\omega_{1}+\omega_{2})\vec{E}(\omega_{1})\cdot\{\partial _{\vec{k}}\rho_{nn}^{(1)}(\omega_{2})+i\sum_{m}[\vec{A}_{mn}(\vec{k})\rho_{mn}^{(1)}(\vec{k},\omega_{2})-\rho_{nm}^{(1)}(\omega_{2})\vec{A}_{nm}(\vec{k})]\}, \\  \nonumber
     \rho_{nm}^{(2)}(\vec{k},\omega_{3}) &=& \frac{-ie}{\hbar\omega_{3}+\epsilon_{nm}}\sum_{\omega_{1},\omega_{2}}\delta(\omega_{3},\omega_{1}+\omega_{2})
     \vec{E}(\omega_{1})\cdot\{\partial_{\vec{k}}\rho_{nm}^{(1)}(\vec{k},\omega_{2})+                             
     i\sum_{l}[\vec{A}_{ln}(\vec{k})\rho_{lm}^{(1)}(\vec{k},\omega_{2})-\rho_{nl}^{(1)}(\vec{k},\omega_{2})\vec{A}_{ml}(\vec{k})]\}, 
   \end{eqnarray}
    \end{widetext}
   respectively.

   \emph{Dc current}   Up to second order response, the intra-band dc current reads,
   %\begin{widetext}
   \begin{equation}\label{intra_part}
    \begin{split}
    J^{intra}_{u}(\omega_{3})=\frac{e}{\hbar}\sum_{n}\int d\vec{k}\partial_{u}\epsilon_{n}(\vec{k})\rho_{nn}^{(2)}(\vec{k},\omega_{3}).
    \end{split}
    \end{equation}
  %  \end{widetext}
   With $\omega_{3}\rightarrow 0$, the ratio d$J^{intra}_{z}$(t)/dt is zero\cite{SM}, i.e., intra-band dc current is vanishing in system with TRIS and initial condition $J^{intra}_{z}$(t=0)=0. The second order inter-band dc current reads,
   \begin{equation}\label{Inter_part}
    \begin{split}
    J^{inter}_{u}(0)=\frac{e^{3}}{\hbar^{2}}\sum_{\alpha,\beta}\sum_{\omega}E_{\alpha}(-\omega)E_{\beta}(\omega)    \\
    \sum_{n,m}\int d\vec{k}\partial_{u}(A_{mn}^{\alpha}(\vec{k})A_{nm}^{\beta}(\vec{k})) \frac{\rho_{mm}^{(0)}(\vec{k})-\rho_{nn}^{(0)}(\vec{k})}{\hbar\omega-\epsilon_{nm}+i\eta}.
    \end{split}
    \end{equation}
 In the case of resonant excitation, $\hbar\omega=\epsilon_{nm}$, the inter-band dc current reads,
 \begin{equation}\label{Inter_part2}
    \begin{split}
    J^{inter}_{u}(0)=\frac{i\pi e^{3}}{\hbar^{2}}\sum_{\alpha,\beta}\sum_{\omega}E_{\alpha}(-\omega)E_{\beta}(\omega)    \\
    \sum_{n,m}\int d\vec{k}\partial_{u}(A_{mn}^{\alpha}(\vec{k})A_{nm}^{\beta}(\vec{k}))(\rho_{mm}^{(0)}(\vec{k})-\rho_{nn}^{(0)}(\vec{k})).
    \end{split}
    \end{equation}
 For linearly polarized light, $E_{\alpha}(-\omega)E_{\beta}(\omega)=E_{\alpha}(\omega)E_{\beta}(-\omega)=\frac{1}{2}|E(\omega)|^{2}$, the second order optical conductivity reads,
   \begin{equation}\label{sigma_L}
    \begin{split}
    \sigma^{(2)}_{u\alpha\beta}(0)=\frac{\pi e^{3}}{\hbar^{2}}    
    \sum_{m}\int d\vec{k}\partial_{u}G_{mm}^{\alpha\beta}(\vec{k})\rho_{mm}^{(0)}(\vec{k}),
    \end{split}
    \end{equation}
 in which $G_{mm}^{\alpha\beta}(\vec{k})=\partial_{\alpha}A_{mm}^{\beta}(\vec{k})+\partial_{\beta}A_{mm}^{\alpha}(\vec{k})$ is the symmetric quantum metric in momentum space. Eq.~(18) demonstrates the LPGE is determined by the integral of gradient of quantum metric in momentum space, and the LPGE usually is not quantized. We noted that, in the low frequency limit, the SHG under linearly polarized light in asymmetric insulator is also related to this symmetric quantum metric\cite{ZL17}. For circularly polarized light, $E_{\alpha}(-\omega)E_{\beta}(\omega)=-E_{\alpha}(\omega)E_{\beta}(-\omega)=i\frac{1}{2}|E(\omega)|^{2}$, the second order optical conductivity reads,
 \begin{equation}\label{sigma_C}
    \begin{split}
    \sigma^{(2)}_{u\alpha\beta}(0)=\frac{\pi e^{3}}{\hbar^{2}}\sum_{\alpha,\beta}    
    \sum_{m}\int d\vec{k}\partial_{u}F_{mm}^{\alpha\beta}(\vec{k})\rho_{mm}^{(0)}(\vec{k}),
    \end{split}
    \end{equation}
  in which $F_{mm}^{\alpha\beta}(\vec{k})=\partial_{\alpha}A_{mm}^{\beta}(\vec{k})-\partial_{\beta}A_{mm}^{\alpha}(\vec{k})$ is the antisymmetric Berry curvature in momentum space. Eq.~(19) demonstrates that the CPGE is determined by the integral of gradient of Berry curvature in momentum space. In the system with spatial inversion symmetry and TRIS, $F_{mm}^{\alpha\beta}(\vec{k})$ is vanishing\cite{Xiao10}, and $A_{mm}^{\beta}(\vec{k})$ is constant. Therefore, both LPGE and CPGE are vanishing if the system has both TRIS and spatial inversion symmetry.

  Eqs.~(18) and (19) are main discoveries of this work, and they demonstrate that LPGE and CPGE have underlying connection with the symmetric and antisymmetric parts of QGT in momentum space, respectively. The nonlinear Hall effect\cite{Fu15}, i.e., the direction of dc current is perpendicular to the plane of electric field, can exist under both linearly and circularly polarized field. Especially, under circularly polarized light, $\sigma^{(2)}_{u\alpha\beta}(0)$ can be approximately quantized in chiral topological semimetal, if only the electron around the $\Gamma$ point is excited by light with low frequency. For example, in the band structure of chiral topological semimetals RhSi and CoSi\cite{DH19,MZH19,Sato19,MZH17,ZSC17,ZL19,TM18}, the Berry curvature accumulated around the $\Gamma$ point is $F_{mm}^{xy}(\vec{k})=\lambda k_{z}/|k|^{3}$, and $\partial_{z}F_{mm}^{xy}(\vec{k})=\lambda/|k|^{3}$, where $\lambda$ is the Chern number of targeted band. Therefore, second order conductivity $\sigma^{(2)}_{zxy}(0)$ is proportional to $\lambda$, and affords a simple approach to detect the Chern number of targeted band if circularly polarized with appropriate frequency is applied.

  In summary, we explored the underlying connection between GPE and QGT. We concluded that the gradient of the symmetric part of QGT is related to the LPGE, while the gradient of antisymmetric part (Berry curvature) is related to the CPGE. Our work afforded an alternative interpretation for PGE in the view of QGT, and classified the underlying connection between CPGE (LPGE) and Berry curvature (quantum metric). The CPGE can be approximately quantized in chiral topological semimetals under circularly polarized with appropriate frequency.

%%%%%%%%%%%%%%%%%%%%%%%%%%%%%%%%%%%%%%%%%%%%%%%%%%%%%%%%%%%%%%%%%%%%%%%%%%%%%%%%%%%%%%%%%%%%%%%%%%%%%%%%%%%%%%%%%%%%%%%%%%%%%%%%%%%%%%%
%\section{Summary}

%%%%%%%%%%%%%%%%%%%%%%%%%%%%%%%%%%%%%%%%%%%%%%%%%%%%%%%%%%%%%%%%%%%%%

%\section{Acknowledgements}
ZL is supported by the National Natural Science Foundation of China (11604068). TI and TT are supported by MEXT via ¡°Exploratory Challenge on Post-K Computer¡± (Frontiers of Basic Science: Challenging the Limits). The calculations were performed on the Hokusai system (Project No. Q20246) of Riken. H.B.S. is grateful for support from the Society of Interdisciplinary Research (SOIREE)and HKUST (IGN17SC04; R9418).
%%%%%%%%%%%%%%%%%%%%%%%%%%%%%%%%%%%%%%%%%%%%%%%%%%%%%%%%%%%%%%%%%%%%%%%
%%\appendix*

%%%\section{Appendixes}
%%% The SHG under external optic field with low frequency limit reads (in atomic unit),
%%% \begin{equation}\label{dipole}
%%%   \begin{aligned}
%%    \chi_{ijj}^{(2)}(2\omega)=& \frac{1}{(2\pi)^{3}}\int\int\int_{-\pi}^{\pi}d^{3}\textbf{k}\{\frac{\langle g|x_{i}|n\rangle\langle n|x_{j}|m\rangle\langle m|x_{j}|g\rangle}{(2\omega-\omega_{ng})(\omega-\omega_{ng})} \\
%%    & +\frac{\langle g|x_{j}|n\rangle\langle n|x_{j}|m\rangle\langle m|x_{i}|g\rangle}{(2\omega+\omega_{ng})(\omega+\omega_{ng})}-\frac{\langle g|x_{j}|n\rangle\langle n|x_{i}|m\rangle\langle m|x_{j}|g\rangle}{(\omega+\omega_{ng})(\omega-\omega_{ng})}}.
%%%  \end{aligned}
%%% \end{equation}
%%% With only two bands, the $\chi^{2}$ reads,

%\nocite{*}


\end{document}